\documentstyle[preprint,aps]{revtex}
\newcommand{\be}{\begin{equation}}
\newcommand{\ee}{\end{equation}}
\newcommand{\bea}{\begin{eqnarray}}
\newcommand{\eea}{\end{eqnarray}}

\newcommand{\p}{\partial}

\tightenlines 

\begin{document}

\draft
\preprint{\small   }

\title{Initial data for gravity \\coupled to scalar, electromagnetic 
 and Yang-Mills fields}

\author{Viqar Husain\footnote{
Email: husain@physics.ubc.ca}}

\address{\baselineskip=1.4em Department of Physics and Astronomy,\\
University of British Columbia,\\
6224 Agricultural Road \\
Vancouver, BC V6T1Z1, Canada}

\maketitle 
 
\begin{abstract}

We give ansatze for solving classically the initial value constraints
of general relativity minimally coupled to a scalar field,
electromagnetism or Yang-Mills theory.  The results include both
time-symmetric and asymmetric data.  The time-asymmetric examples are
used to test Penrose's cosmic censorship inequality. We find that the
inequality can be violated if only the weak energy condition holds.
  
\end{abstract}
\bigskip

\section{Introduction}

The problem of finding solutions of the initial value constraints
(``initial data'') of general relativity in vacuum, or coupled to
matter, is of interest from both mathematical and physical
viewpoints \cite{books}. The mathematical questions concern existence
and uniqueness of solutions for various classes of initial data and
topology of Cauchy surfaces. The more physical questions concern
cosmic censorship, and numerical evolution of physically relevant
initial data. 

There has been much work on construction of initial data sets 
using analytical and numerical techniques, as well as combinations 
of both. Analytical methods initiated by Lichnerowitz, and  
further developed by York, are now standard material \cite{smarr}. 
Combinations of analytic and numerical techniques continue to be 
investigated, particularly for problems of physical interest. 

In vacuum, perhaps the simplest initial data is the spherically
symmetric ``throat'' connecting two asymptotically flat regions, which
has a generalization to $N$ throats with masses and charges \cite{mw}.
There is also a wormhole solution due to Misner \cite{misnerwh}, where
the wormhole is a handle on flat space. Other multi-black hole 
solutions, having time and inversion symmetry, have also been derived 
\cite{misner2}, with subsequent generalization by Bowen and York \cite{by}. 
In more recent work, a variation on the theme of analytic-numerical 
data for $N$ black holes with arbitrary momenta and spin numerically 
is given in \cite{bb}.

Most of the analytic solutions of the initial data problem are in
vacuum. Analytic solutions with specific matter coupling are less 
studied. Such solutions are potentially useful,  both for their intrinsic 
value, and for providing  starting points for numerical time evolution 
schemes. As an example of the latter, analytic initial data for scalar 
field coupling  may be  useful for further study of the spherically 
symmetric collapse problem \cite{chop}, which has been a subject of 
recent interest. This collapse problem has also been studied for matter 
couplings other than the scalar field \cite{cars}, but always with 
numerically generated initial data. 

A second reason for seeking analytic initial data sets is their 
potential usefulness for probing the cosmic censorship conjecture. 
One statement of this conjecture is that trapped surfaces in a spacetime 
always lie inside event horizons. This has led Penrose to suggest an 
``initial data test'' for weak cosmic censorship \cite{pen3}. The test 
is an inequality relating the area of the outermost trapped surface 
$A(S)$, the area of the event horizon $A_{EH}$, and the ADM mass M of 
asymptotically flat initial data, all for matter satisfying reasonable 
energy conditions. The inequality is 
\be 
       A(S)\le A_0\le A_{EH}\le 16\pi M^2,
\label{ineq}
\ee     
where $A_0$ is the area of a surface that encloses the outermost 
marginally trapped surface $S$. The first and second inequalities are 
reasonable from a physical viewpoint because, as matter collapses the 
region containing trapped surfaces gets larger, and in the long time 
static (or stationary) limit $A(S)\rightarrow A_{EH}$. The third 
inequality allows the possibility 
that all the mass does not end up in the black hole. The content of the 
cosmic censorship test is this: If initial data exists such that the area 
of the outermost trapped surface is larger than $16\pi$ times the 
square of the ADM mass of the data, then weak cosmic censorship is 
violated. 
 
Trapped surfaces may be present on an initial data surface for certain
ranges of the initial data parameters. The boundary separating trapped
and untrapped regions on the initial data surface is the apparent
horizon.\footnote{There is also an alternative terminology in the 
literature: only the {\it outermost} marginally trapped surface is called 
the apparent horizon. Here we refer to all marginally trapped surfaces 
as apparent horizons, as in Ref. \cite{hawk-ell}.}  
  On this horizon the outward expansion of light rays 
vanishes. In terms of initial data, a closed spatial 2-surface $S$ with 
normal
$s^a$ is {\it outer trapped} if
\be 
(q^{ab} - s^as^b)(K_{ab} + D_as_b)=0,
\label{aheqn}
\ee 
where $K_{ab}$ is the extrinsic curvature of the initial data surface,
and $q_{ab}$ is its metric. \footnote{In terms of the gravitational 
momentum $\tilde{\pi}^{ab}=\sqrt{q}(K^{ab}- K q^{ab})$, where 
$K=K^{ab}q_{ab}$, this equation is $ \tilde{\pi}^{ab}s_as_b = 
\sqrt{q} D_as^a. $} This equation expresses the
vanishing on $S$ of the outward null expansion.\footnote{The vanishing
of the inward null expansion is obtained by changing the sign of $s^a$
in (\ref{aheqn}). This corresponds to a white hole situation.}

Eqn.  (\ref{ineq}) may be viewed as  providing a relation between local 
and global information in the sense that the apparent horizon is a 
solution of a differential equation involving the (local) phase space 
variables, whereas the ADM mass is an integral of a phase space function 
over the boundary of the initial data surface. While it is true that 
the apparent horizon is also an embedded 2-surface (and hence `global'), 
its shape and size are subject to change due to local matter flows. 
 
There have been a number of successful tests of the inequality
(\ref{ineq}). The first were for null shells with flat interior, in
certain special cases \cite{pen3,gibb1}. However, it has been shown
recently that the inequality holds generally for null shells with flat
interior \cite{trud}, (see also \cite{gibb2}).  For arbitrary matter
satisfying the dominant energy condition, it was proven by Jang and
Wald \cite{jw} that time-symmetric initial data (zero extrinsic
curvature) satisfies this inequality. The status of the inequality for
the time-asymmetric cases is in general open (with the exception 
of null shells with flat interior mentioned above\cite{rev1,rev2}). 

With the above motivation, it is the purpose of this paper to present 
classes of analytic initial data sets for certain matter couplings, 
discuss some of their properties, and use them as probes of Penrose's 
inequality. 

The matter couplings discussed are to the massless scalar field,  
electromagnetism, or Yang-Mills theory. It is 
based on the conformally flat ansatz for the spatial metric
\cite{books}. The next section gives solutions for a class of
time-symmetric cases, followed in Section III by time-asymmetric data
for scalar field coupling. This section also contains results
pertaining to Penrose's inequality for time-asymmetric data: the
inequality  is satisfied for all the examples considered, with the
exception of cases where only the weak energy condition holds.

\section{Time-symmetric data}
 
Consider Einstein gravity minimally coupled to a massless scalar
field with no self interaction. In the Hamiltonian formulation, the 
phase space variables are the canonically conjugate pairs 
$(\phi,\tilde{P})$ for the scalar field, and $(q_{ab}, \tilde{\pi}^{ab})$ 
for the gravitational field. The initial value constraints are
\bea 
{\cal H} &\equiv& {1\over \sqrt{q}} 
               G_{abcd}\tilde{\pi}^{ab}\tilde{\pi}^{cd} 
              - \sqrt{q}^{(3)} R 
              +  {1\over \sqrt{q}}\tilde{P}^2 
              + \sqrt{q} q^{ab} \p_a\phi \p_b \phi = 0, \label{Ham}\\
{\cal C}^a  &\equiv& \p_b\tilde{\pi}^{ba} + \tilde{P}q^{ab}\p_b \phi=0.    
\eea
where $G_{abcd} = (g_{ac}g_{bd} + g_{ad}g_{bc} - g_{ab}g_{cd})/2$ is 
the DeWitt supermetric, $D_a$ is the covariant derivative of 
$q_{ab}$, and $\tilde{\ }$ denotes densities of weight one.   

Consider the topology of the spatial slice $\Sigma$ to be $R^3$, with
the rectangular coordinates $x,y,z$. The usual time-symmetric 
conformal ansatz for the vacuum constraints is
\be 
\tilde{\pi}^{ab} = 0\ ; \ \ \ \ \ \ q_{ab} = \psi^4 \delta_{ab},
\ee
where $\delta_{ab}$ is the flat Euclidean 3-metric. With this ansatz 
the diffeomorphism constraint is identically satsified and the scalar 
constraint ${\cal H}$ becomes the Laplace equation. The simplest one 
throat (or Schwarzschild) solution of the Hamiltonian constraint is 
\be
\psi = 1 + {m \over 2r}, 
\label{sch}
\ee 
where $m$ is a constant (the ADM mass) and $r^2 = x^2+y^2+z^2$. 
To avoid the singular behavior at $r=0$ this solution is viewed 
as being on $R^3 -\{ 0\}$. Evolution 
of this data gives the Schwarzschild solution. This result generalizes 
to $N$ black holes \cite{mw} with masses $m_i$ at the coordinate 
locations $r_i$  on $R^3 - \{ \rm{ points\ excised\ for\ each\ 
black\ hole}\}$:
\be 
\psi = 1 + \sum_{i=1}^{N} {m_i\over 2|r-r_i|}. 
\ee

Consider now the following ansatz for the Einstein-scalar field 
theory.  Again with the spatial slice $\Sigma \approx R^3$, set 
\be 
\tilde{P}=0\ , \ \ \ \ \ \tilde{\pi}^{ab}=0\ ,\ \ \ \ \  
q_{ab} = \psi^4 \delta_{ab}\ , 
\ee 
with arbitrary initial scalar field $\phi(x,y,z,t=0)$. The 
Ricci scalar of $q_{ab}$ is 
\be
^{(3)}R = -8\nabla^2\psi/\psi^5.
\ee 
With this, the Hamiltonian constraint (\ref{Ham}) 
becomes   
\be 
\left( \nabla^2 + {1\over 8}\ \delta^{ab} \p_a\phi\p_b\phi \right)\ 
\psi(x,y,z) = 0. 
\label{fsc}
\ee 
This is the Schr\"odinger equation in three dimensions with 
potential $V(x,y,z)$ and eigenvalue $E$ related by 
\be
{2m\over \hbar^2} [E - V(x,y,z)] 
= {1\over 8}\ \delta^{ab}\ \p_a\phi\p_b\phi. 
\label{fscb}
\ee
Since the r.h.s. is positive or zero, the physical solutions of the
Hamiltonian constraint must satisfy $E-V \ge 0$. This leads to an
infinite class of solutions, derived from any $V$ and $E$, if the scalar 
field is allowed to be arbitrary, and possibly singular on the initial 
data slice. However, since regular initial data is more physically 
relevant,  the ``wavefunctions'' $\psi$ of interest, for example for 
matter collapse, will mainly be those which do not vanish anywhere. 
This criteria appears to exclude normalizable wavefunctions, where 
$\psi$ vanishes somewhere.  A further restriction is that the scalar 
field must be real; there are $E$ and $V$ for which this is not the 
case for all $r$. These conditions are however not  drawbacks to the 
analogy with the Schrodinger equation, as it is still possible to remove 
the normalizability boundary condition to get many instances of 
non-singular initial data sets with real $\phi$, as we see below.

(For a free massless scalar field, the Hamiltonian constraint(\ref{fsc})
can also be converted into the Poisson equation $\nabla^2 \psi =
-V_aV_b \delta^{ab}$, where the vector field $V_a$ is freely specified; 
the  initial scalar field is determined from any solution $\psi$
using $\p_a\phi = V_a/\sqrt{\psi}$. Also, neither this nor the previous 
case  gives a linear equation for $\psi$  if the scalar field is massive 
or has an arbitrary potential term.)  

It is straightforward to obtain some simple explicit examples of
regular solutions (ie. nowhere vanishing $\psi$), and a physical 
picture.

\noindent (i) {\it Solutions in spherical symmetry}: Given a radial 
wavefunction $\psi(r)$ with eigenvalue $E$ for a potential $V(r)$, the 
initial scalar field $\phi(r,t=0)$ from (\ref{fscb}) is
\be 
\phi(r,t=0) = \sqrt{{16 m\over \hbar^2}}\int^r dr'\ \sqrt{E-V(r')}.
\ee
Thus, the exact initial scalar field which solves the Hamiltonian 
constraint is proportional to phase of the WKB wavefunction for any 
potential $V(r)$ and energy $E$, with the restriction that 
$E-V(r)\ge 0$ to get real $\phi(r,t=0)$.   

A particular solution in spherical symmetry: consider 
the scalar field pulse defined by 
\be 
\phi(r,t=0)  = \cases{ 0  &if $0\le r<r_0$ \cr
                    \sqrt{2}\ {\rm ln}r    &if $ r_0\le r \le r_1 $ \cr 
                           0    &if $  r >r_1$ 
                          }
\label{data1}
\ee
where $r_0,r_1$ are parameters giving the pulse width. One solution 
for $\psi(r)$ is 
\be  
\psi(r,t=0)  = \cases{ 1  &if $0<r<r_0$ \cr
                    A/\sqrt{r} + B\ {\rm ln}r/\sqrt{r}   
                             &if $ r_0\le r \le r_1 $ \cr 
                    C + D/2r  &if $  r >r_1$ \ .
                   }
\label{data1sol}
\ee
Here we have chosen the 3-metric to be  flat in the ``inner'' region, 
Schwarzschild like in the ``outer'' region, and determined by the 
scalar field shell in the middle. The coefficients $A$, $B$, $C$ 
and $D$ are determined by continuity of $\psi(r)$ and its first 
derivative at the interfaces $r=r_0$ and $r=r_1$ to be
\be 
A = \sqrt{r_0}\left( 1 - {1\over 2}\ {\rm ln}r_0 \right) ; 
\ \ \ \ \ \ \ \ 
B =  {\sqrt{r_0}\over 2}.
\ee
\be 
C= {1\over \sqrt{\alpha}}
\left[1 - {1\over 4}\ {\rm ln}\alpha \right],\ \ \ \ \ \ \ 
D = r_0 {\sqrt{\alpha}\over 2}\ {\rm ln}\alpha,\ 
\ee
where $\alpha \equiv r_1/ r_0 >0$ gives the ``width'' of the pulse.
The ADM mass of this solution $M = CD$ is a function of the
``distance'' $r_0$ of the pulse from the origin and the number
$\alpha$.  It is possible to have a non-flat vacuum in the inner
region by setting $\psi = 1+m/2r$ in the inner region. The data then
has the additional parameter $m$, and is singular.
 
In spherical symmetry with vanishing extrinsic curvature, the apparent 
horizon equation (\ref{aheqn}) simplifies to $g^{ab}\p_a R\p_b R = 0\ $,
where $R=r\psi^2(r)$ is the radial coordinate of the nested spheres and 
$g^{ab}$ is the spacetime metric. For the conformally flat ansatz for the 
3-metric, this leads to the equation   
\be
\psi + 2r \psi'=0.
\label{ahsp}
\ee
Thus, apparent horizons exist on the initial data surface if there are 
real solutions $r=r_{AH}$ of this equation, with $r_{AH}$ lying in the 
appropriate region(s) for each $\psi$. For the solution (\ref{data1sol}), 
there are {\it no} apparent horizons in the region of non-vanishing 
scalar field because (\ref{ahsp}) reduces to 
$ 2B/\sqrt{r} = 0$. 

There are other examples where this is not the case, and
apparent horizons are present. One interesting case is a solution
which is asymptotically flat {\it without} matching to Schwarzschild
exterior.  Consider
\be 
\phi = {\sqrt{2}C\over r},
\label{1/r}
\ee
where $C$ is a constant. 
The corresponding solution of the Hamiltonian constraint is 
\be 
\psi = A\ {\rm cos}\left( {C\over 2r} \right) 
    +  B\ {\rm sin}\left( {C\over 2r} \right).  
\ee
This solution is asymptotically flat. For large $r$,
the constant $A$ is a conformal factor which may be set to unity.
Comparing with Schwarzshild shows that the ADM mass $M$ of this data
(with $A=1$) is $M = BC$. It has a curvature singularity at $R=r\psi^2
= 0$. It is interesting to note that there are an infinite number of
zeroes of $\psi$ and a corresponding number of apparent horizons near
these zeroes, even though the scalar field is singular only at $r=0$.
Surprisingly, for this data there are regions of parameter space 
where Penrose's inequality appears to be violated. One such region is at 
and near the point $B=0.01$ and $C=2.00$ (with $A=1$), where 
$2M - R_{AH}=-0.286$. However, this does not provide a counterexample 
of cosmic censorship because the conformal factor $\psi$ goes to zero 
outside the horizon \cite{tedcc}.  We note that Penrose's inequality 
may still be tested, and is satisfied, for the apparent horizon that 
lies in the region connected to spatial infinity; ie. as one comes in 
from infinity, the horizon is encountered  before any zero of $\psi$.  

It is possible to get solutions with a finite number of apparent
horizons by patching flat space in the inner region, from $r=0$ to
some $r=r_1$.  Set $r_1=1$, and $\phi=0$ and $\psi=1$ for $0\le r <
1$. Then, with $\phi$ as in (\ref{1/r}) for $r\ge 1$, the solution is
\be 
\psi(r) = A(C)\left[ {\rm cos} \left(C\over 2r\right) 
+ {\rm tan}\left({C\over 2}\right) {\rm sin}\left({C\over  2r}\right)
\right]\ \ \ \ \ \ \ {\rm for}\  r \ge 1,
\ee 
where 
\be 
A(C) = [ {\rm cos}(C/ 2) + {\rm tan}(C/ 2)\ {\rm sin}
(C/2) ]^{-1}.
\ee 
The ADM mass of this one parameter $(C)$ configuration 
is 
\be 
  M = A^2(C)\ C\ {\rm tan}\left( {C\over 2} \right). 
\ee
Now, depending on the value of $C$, there are zero or a 
finite number of apparent horizons in the matter region. 
It is also possible to construct a solution with a finite 
number of horizons by patching Schwarzschild rather than flat 
data in the inner region. It is again possible to choose 
parameters such that a horizon is encountered before 
a zero of $\psi$, as $r$ is varied inward from infinity.  
 
\noindent (ii) {\it Scalar field wormhole:} Away from spherical 
symmetry, the situation is virtually unchanged. Consider perturbing 
Misner's wormhole solution \cite{misnerwh} by the presence of a compact 
scalar field pulse. The mathematical problem is one of matching 
an `interior' vacuum solution to a pulse solution, which in turn is 
matched to an `exterior' vacuum solution. (Vacuum solution means a 
solution of the constraint equations without matter sources.) The former 
and latter are just Misner's $\psi$, with different parameters in each 
region to account for the mass in the scalar field. The solution in the 
intermediate region depends on the scalar field pulse, which determines 
the potential in the Schrodinger equation analogy. The details are a 
matching problem not unlike the one in spherical symmetry considered 
above. Evolution of such initial data would be of interest for 
seeing how presence of matter affects the gravitational radiation from  
the two black hole problem in the so called ``close'' limit \cite{pp}.

 {\it Electromagnetism:} There are similar results for coupling 
to electromagnetism, with phase space variables $(A_a, \tilde{E}^a)$ 
and Hamiltonian density $(\tilde{E}^a\tilde{E}^b 
+ \tilde{B}^a\tilde{B}^b)q_{ab}/2\sqrt{q}$, 
($\tilde{B}^a =\tilde{\epsilon}^{abc}\p_bA_c$). Using the same 
conformal ansatz for the metric and  conjugate momentum, 
$q_{ab}=\psi^4\delta_{ab}$ and $\tilde{\pi}^{ab}=0$, set 
\be
\tilde{E}^a=\psi^2 \tilde{e}^a \ \ \ \ \ \ 
\tilde{B}^a=\psi^2 \tilde{b}^a 
\label{em1}
\ee
where $\tilde{e}^a(x,y,z)$ and $\tilde{b}^a(x,y,z)$ are arbitrarily 
specified. Then the Hamiltonian constraintbecomes 
\be
 \left[\ \nabla^2 + {1\over 16} (\tilde{e}^a \tilde{e}^b\delta_{ab} + 
\tilde{b}^a \tilde{b}^b\delta_{ab}) \ \right]\psi = 0,
\label{emH1}
 \ee
identical in form to the scalar field coupling case (\ref{fsc}).
Thus, given the freely specified fields $\tilde{e}^a$ and
$\tilde{b}^a$, one solves the Hamiltonian constraint for $\psi$, and
then determines the physical electric and magnetic fields via
(\ref{em1}). Note that this ansatz is not the same as the one given by
Misner and Wheeler \cite{mw}, which gives charged black hole data;
their metric ansatz is $q_{ab}=(\psi^2-\chi^2)^2\delta_{ab}$, from
which the scalar constraint gives Laplace equations for $\psi$ and
$\chi$.

As for the scalar field case, there is an alternative specification of
the electric and magnetic fields which converts the Hamiltonian
constraint into the Poisson equation with source given implicitly by
these fields.  This arises by setting
\be 
\tilde{E}^a=\psi^{3/2} \tilde{e}^a \ \ \ \ \ \ 
\tilde{B}^a=\psi^{3/2} \tilde{b}^a,
\label{em2}
\ee 
and gives  
\be
 \nabla^2 \psi = - {1\over 16} (\tilde{e}^a \tilde{e}^b\delta_{ab} +   
\tilde{b}^a \tilde{b}^b\delta_{ab}) 
\label{emH2}
\ee
for the Hamiltonian constraint.  

Of course, neither (\ref{em1}) or (\ref{em2}) solve the diffeomorphism 
constraint unless either the electric field or vector potential $A_a$
is set to zero. Furthermore, there is also the Gauss law for the
electric field. One obvious solution to all the initial value
constraints is obtained by setting $\tilde{E}^a =0$, and solving the
scalar constraint (\ref{emH1}) or (\ref{emH2}) for purely magnetic
initial matter. In this context, one can again consider special cases,
such as spherical or toroidal symmetry, or a wormhole.

Purely electric solutions, other than the monopole, are more difficult
to find without imposing further symmetries, since it is difficult to
simultaneously solve the vacuum Gauss law constraint {\it and} obtain
a linear equation for the scalar constraint. However, this is possible 
if symmetries are imposed. As an example consider the case of 
one translational symmetry and seek solutions $\psi=\psi(x,y)$. For two
arbitrary functions $u(x,y)$ and $v(x,y)$, set \cite{nr}
\be 
\tilde{e}^a = \tilde{\epsilon}^{abc} \p_b u \p_c v ,  
\label{gauss}
\ee
where $\tilde{\epsilon}^{abc}$ is the metric independent Levi-Civita 
tensor density. Then the electric field given by (\ref{em1}) satisfies 
Gauss's law
\be 
\p_a\tilde{E}^a = \p_a(\psi^2\epsilon^{abc}\p_bu\p_cv) = 
2\psi\epsilon^{abc}\p_a\psi\p_bu\p_cv = 0 
\ee
because $u$, $v$ and $\psi$ depend only on two coordinates.  The
electric field lines are tangent to the curves defined by the
intersection of the surfaces $u =$ constant and $v =$ constant. In the
present case, the only non-vanishing component of $E^a$ is $E^z(x,y)$.

Finally, we point out a class of solutions for which both initial 
electric {\it and} magnetic fields are non-zero. This is again in the 
context of one translational symmetry.  Given arbitrary 
$u(x,y)$,  $v(x,y)$ and $b^a(x,y)$, define $e^a(x,y)$ according to 
(\ref{gauss}).  As before, since the solutions of the Hamiltonian 
constraint are  $\psi=\psi(x,y)$, Gauss's law is automatically  
satisfied. The diffeomorphism constraint
\be 
\p_a\tilde{\pi}^{ab} + \tilde{E}^b(\p_a A_b - \p_b A_a) = 0 
\ee
 reduces to 
\be 
\psi^2\tilde{\epsilon}_{abc} \tilde{E}^b \tilde{B}^c = 0. 
\ee
Now, since $\tilde{E}^a$ has only a $z-$component, restricting 
$\tilde{B}^a$ to have only a $z-$ component solves this constraint. 
Furthermore, $\p_a\tilde{B}^a = \p_z(\psi^2(x,y)\tilde{b}^z(x,y))=0$ 
so there are no magnetic sources.  

This procedure gives solutions of all the initial value constraints
with electromagnetic coupling; the data is characterized by two arbitrary
functions of two coordinates.  As there are no electromagnetic
sources, this class of solutions may be called ``geon'' data. This 
result has extensions to any 3-space with one Killing symmetry.

{\it Yang-Mills theory:}  The phase space variables are 
$(A_a^i, \tilde{E}^{ai})$, where $i$ ($ = 1\cdots N^2-1$) 
is the  SU(N)) Lie algebra index. The Yang-Mills contribution to the 
Hamiltonian constraint is 
\be 
H = {1\over 2\sqrt{q}}(\tilde{E}^{ai}\tilde{E}^{b^j} 
    + \tilde{B}^{ai}\tilde{B}^{bj})q_{ab}k_{ij}\ ,
\ee 
where $k_{ij}$ is the Cartan metric, and $\tilde{B}^{ai}= 
\tilde{\epsilon}^{abc}F_{bc}^i$. 
($F_{ab}^i = \p_a A_b^i - \p_b A_a^i + C^i_{\ jk}A_a^jA_b^k$ and  
$C^i_{\ jk}$ are the structure constants of the gauge group). 
The ansatz that reduces the Hamiltonian constraint to the Schrodinger 
equation is similar to (\ref{em1}), namely  
\be
\tilde{E}^{ai}=\psi^2 e^{ai} \ \ \ \ \ \ \tilde{B}^{ai}=\psi^2 b^{ai}.
\label{ym1}
\ee
Thus, given arbitrary $\tilde{e}^{ai}$ and $\tilde{b}^{ai}$, finding 
the conformal factor $\psi$ is again an elementary problem. However, we 
must also find solutions of the diffeomorphism constraint and the 
non-abelian Gauss law 
\be
D_a \tilde{E}^{ai}\equiv\p_a \tilde{E}^{ai} 
+ C^i_{\ jk} A_a^i\tilde{E}^{ak}=0.
\label{ymG}
\ee

One class of solutions of all the constraints is obtained by starting 
with $A_a^i=0$. This immediately solves the diffeomorphism
constraint (because $\tilde{\pi}^{ab}=0$), and the Gauss law becomes 
$\p_a (\psi^2 e^{ai})=0$. Now set
\be 
\tilde{e}^{ai} = C^i_{\ jk} \tilde{\epsilon}^{abc}\p_b u^j\p_c v^k
\ee
where $u^i$ and $v^i$ are $2(N^2-1)$ arbitrary functions. This 
form does give a ``full'' solution of the Gauss law for this case (ie. 
with $A_a^i=0$). This is because its solutions, without any redundancy, 
are parametrized by $2(N^2-1)$ functions for $SU(N)$, since there are 
$N^2-1$ constraints for the $3(N^2-1)$ $\tilde{E}^{ai}$ fields in three 
spatial dimensions. With this ansatz Gauss law becomes
\be 
\p_a(\psi^2 \tilde{e}^{ai}) = 
2\psi C^i_{\ jk} \tilde{\epsilon}^{abc}\p_a\psi\p_b u^j\p_c v^k. 
\ee
 Then, as for the abelian case, there are solutions to 
all the constraints if the problem is reduced so that all variables 
depend on two coordinates. 
 
It is more difficult to find solutions for which both electric 
and magnetic fields are non-vanishing. This is due to the term 
quadratic in the phase space variables in the Gauss law. To see this,  
suppose we give an ansatz specified by an arbitrary gauge field $a_a^i$ 
whose magnetic field is $\tilde{b}^{ai}$, and with $\tilde{e}^{ai}$ 
given by 
\be  
\tilde{e}^{ai}({\bf x}) = \sum_{\bf a} \tilde{\epsilon}^{abc}
\p_b u^{\bf a}\p_c v^{\bf a}\ {\rm Tr}
[ U[a_a,\gamma(u^{\bf a},v^{\bf a})]\tau^i]({\bf x})\ ,
\label{ymg}
\ee
where  $u^{\bf a}$ and $v^{\bf a}$ are a set of 
scalars (labelled by indices ${\bf a,b,}$), $\tau^i$ are matrix 
generators of the group, and  
\be 
U[A_a,\gamma(u,v)]({\bf x}) = 
{\rm P exp}\int_\gamma ds A_a^i(\gamma(s))\tau^i
\ee  
is the holonomy of the gauge field along the loop 
$\gamma(u^{\bf a},v^{\bf a})$, with base point ${\bf x}$, determined by 
the intersection of surfaces $u^{\bf a}=$ constant and $v^{\bf a}=$ 
constant. This is a non-abelian generalization of (\ref{gauss}) \cite{nr}. 
It is straightforward to check that it satisfies the Gauss law for the 
fields $(e^{ai}, a_a^i)$. However, because of the conformal factor $\psi$ 
in (\ref{ym1}), it does not satisfy the Gauss law (\ref{ymG}) for 
the physical fields $(E^{ai}, A_a^i)$. Indeed, from a solution $\psi$ 
of the scalar constraint obtained via (\ref{ym1}), we find for 
(\ref{ymg}) that 
\be 
D_a\tilde{E}^{ai}=D_a(\psi^2\tilde{e}^{ai})
= \psi^2 f^i_{\ jk}(A_a^j - a_a^j)\tilde{e}^{ak} 
+ 2\psi \tilde{e}^{ai}\p_a\psi\ \ne 0,
\label{gl2}
\ee
where $A_a^i$ is a gauge field associated with the physical magnetic 
field $\tilde{B}^{ai}=\psi^2 \tilde{b}^{ai}$. 

One way to solve (\ref{gl2}) is to assume, at the outset, that
$e^{ai}$ and $a_a^i$ are parallel in the internal direction, 
ie. that $\tilde{e}^{ai} = \tilde{e}^at^i$ and $a_a^i = a_at^i$ 
for a fixed Lie algebra vector $t^i$. This means that $[a_a,a_b]=0$,  
and $a_a^i$ and $A_a^i(\psi,a_a^i)$ are also parallel in the internal 
direction. With these conditions on the initial configuration, the 
first term in (\ref{gl2}) vanishes. The second term vanishes if we
assume, as before, that all fields depend on only two of the spatial
coordinates $(x,y)$, ie. that there is a translation Killing 
symmetry. This means that the YM electric field from (\ref{ymg}) has 
only a $z-$component. 

The diffeomorphism constraint 
\be 
\epsilon_{abc}\tilde{E}^{bi}\tilde{B}^{cj}k_{ij} = 0 
\ee
is satisfied if the YM magnetic field is also restricted to have 
only a $z-$ component.  Finally, note that $D_a\tilde{B}^{ai} = 
\p_z\tilde{B}^{zi} + [A_z,\tilde{B}^z]^i = 0$ because $B^{zi}$ is a 
function only of $x,y$, and the commutator vanishes by the ansatz used 
to solve the Gauss law. Thus there are no magnetic sources.   
 
In this way one can obtain solutions to all the constraints for the
time symmetric Yang-Mills problem, with the conformal ansatz for the
metric. Such solutions are parametrized by functions of two variables
$u^{\bf a}$ and $v^{\bf a}$. Because of the way the Gauss law is
solved, the Yang-Mills matter for this data are arranged into
``non-abelian lines of force.'' This data is therefore a direct 
generalization of the electromagnetic data given above.
 
{\it Charged scalar field}: For gravitational coupling to both 
electromagnetism and a charged scalar field, the Hamiltonian 
constraint has the additional term  
\be 
\sqrt{q} q^{ab} {\cal D}_a\phi {\cal D}_b\phi^* 
+ \tilde{P}\tilde{P}^*/2\sqrt{q}, 
\ee
where ${\cal D}_a = \p_a + A_a$ and * denotes complex conjugation. 
With the time-symmetric conformal ansatz, and $\tilde{P}=0=A_a$, the 
Hamiltonian constraint is  
\be 
8 \nabla^2\psi + \left( \delta^{ab}\p_a\phi \p_b\phi^* 
+ \tilde{e}^a\tilde{e}^b \delta_{ab} \right) \psi = 0,  
\label{Hchsc}
\ee 
which must be solved with the Gauss law 
$\p_a\tilde{E}^a = \tilde{P}\phi^* + \tilde{P}^*\phi$.  

A  large class of solutions of this system can be obtained when 
there is one translational symmetry. As before set $\tilde{e}^a
=\tilde{\epsilon}^{abc} \p_bu\p_cv$, where $u$ and $v$ are arbitrary
functions of the coordinates $x,y,z$.  $\phi$ is also
arbitrarily. Then the linear equation (\ref{Hchsc}) can be solved for
$\psi$. With a solution $\psi$, the Gauss law for $\tilde{E}^a =
\psi^2\tilde{e}^a$ (with $\tilde{P}=0$) becomes
\be 
\tilde{e}^a \p_a\psi =  
\tilde{\epsilon}^{abc} \p_bu \p_cv \p_a \psi = 0.
\ee 
Thus, if all the functions $u,v,\psi$ depend on only two coordinates, 
the Gauss law is identically solved. 

\section{Time-asymmetric data} 

So far we have restricted attention to time-symmetric
(ie. $\tilde{\pi}^{ab}=0$) initial conditions. We now consider some
cases which relax this condition, but still restrict attention to the
conformally flat form of the three-metric. The goal of the ansatze is
to obtain more general initial data by attempting to convert the
Hamiltonian constraint into a solvable equation. This is useful not
only for obtaining more general classes of solutions, but also for
addressing the cosmic censorship conjecture via the initial data test
(\ref{ineq}); as noted in the introduction, the status of this initial
data test is open for the general time-asymmetric case.

If $\tilde{\pi}^{ab}\ne 0$, the vector (or spatial diffeomorphism)
constraint must also be solved. It is obvious (and well known) that in
vacuum an easy way to solve this constraint is to set
$\pi^{ab}=Aq^{ab}$, where $A$ is a constant. When matter is present,
it is possible to obtain initial data by generalizations of this, as
we now describe.

Consider the scalar field case with the spherically symmetric 
ansatz   
\be 
q_{ab} = \psi^4(r)\delta_{ab}  \ \ \ \ \ \ 
\tilde{\pi}^{ab} = \psi^{-4} \tilde{P}(r)\left( 
\alpha  n^an^b  + \beta \delta^{ab} \right),
\label{asym1}
\ee
where $\tilde{P}(r)$ is the scalar field momentum density, 
$n^a = x^a/r$ is the unit radial vector, and $\alpha$ and $\beta$  
are constants. The diffeomorphism and Hamiltonian constraint become
\be
 (\alpha + \beta)\p_r(\psi^{-4}\tilde{P}) 
+ {2\alpha \psi^{-4}\tilde{P}\over r} +  
\psi^{-4}\tilde{P}\p_r \phi=0,       
\ee
\be
\left( {\alpha^2\over 2} - \alpha \beta - {3\beta^2\over 2} + 1  \right) 
\psi^{-6} \tilde{P}^2  
+ 8\psi\nabla^2\psi + \psi^2 \p_a\phi\p_b\phi\delta^{ab} = 0.  
\label{H0}
\ee
The diffeomorphism constraint gives 
\be  
 \tilde{P} = {C\psi^4\over r^{2\alpha/\alpha + \beta}} 
{\rm exp}\left[ -{\phi\over \alpha + \beta} \right]. 
\label{P1}
\ee 
One class of solutions is obtained by arranging cancellation of the 
gravitational and scalar momenta. Thus, with $\tilde{P}$ arbitrary,  
set the coefficient of $\tilde{P}^2$ to zero. This gives 
\be 
\alpha = \beta \left( 1 \pm 2\sqrt{ 1 - {1\over 2\beta^2}} \right). 
\label{cond}
\ee
Now, if the scalar field is chosen to fall off sufficiently fast at
large $r$, $q_{ab}$ falls off like Schwarzshild at spatial infinity.
However, from (\ref{cond}) it is evident that the leading order term
of $\tilde{\pi}^{ab}$ cannot be $O(1/r^2)$; the falloff is slower than
this. For the special case $\psi=1$ (flat slice), the required falloff 
is $r\sim r^{-3/2}$ (see below). Even this violates (\ref{cond}).
This means that this data is not asymptotically flat. 

There is a generalization of this data  
obtained by requiring that $\p_r\phi$ be proportional 
to $\tilde{P}$, again with $\psi=1$ in (\ref{H0}). This requires 
$\phi\sim {\rm ln}r$, which means that $\tilde{P}\sim r^{-1}$. 
Therefore this also does not give asymtotically flat data.   

In order to find asymptotically flat data, let us consider the general 
spherically symmetric ansatz with flat spatial metric,\footnote{I thank 
Ted Jacobson for suggesting the use of flat slices.} ie. $\psi = 1$, or 
$q_{ab} = \delta_{ab}$.
 The general form of $\tilde{\pi}^{ab}$ is 
\be 
\tilde{\pi}^{ab} = f(r) n^an^b + g(r) \delta^{ab}.
\label{gf}
\ee
The diffeomorphism constraint, with matter current $J^a=0$, is 
\be 
 f(r) = -{1\over r^2} \int dr\ (r^2\p_rg) + {c\over r^2},
\label{asymdiff}
\ee
($c=$ constant) and the Hamiltonian constraint is 
\be 
\tilde{\pi}^{ab}\tilde{\pi}_{ab} - {1\over 2}(\tilde{\pi}^a_a)^2 
+ \rho 
= {1\over 2}(f + g) (f - 3g) + \rho = 0, 
\label{asymH}
\ee
where $\rho$ is the matter energy density. 

In vacuum ($\rho=0$), the solution of the constraint equations 
may be found by taking $g=r^\alpha$ in the above equations. The 
solution is 
 \be 
\tilde{\pi}^{ab} =  { C\over r^{3/2}}\ 
(3 n^an^b + \delta^{ab}),
\label{flsch}
\ee
where $C$ is a constant. Since this is vacuum data is spherical
symmetry, it must be data for the Schwarzschild metric. Its mass $M$
may be determined in terms of $C$ by locating the horizon on the
initial data surface. With $s^a = n^a$ in the apparent horizon eqn.
(\ref{aheqn}), and the horizon radius fixed to be $2M$,
eqn. (\ref{aheqn}) gives
\be 
C = \sqrt{M/2}.
\ee 
The corresponding space time metric is \cite{flat}
\be 
ds^2 = -dt^2 + \left( dr + \sqrt{{2M\over r}}dt\right)^2 
+ r^2 d\Omega^2.
\ee

The flat slice data for Schwarzschild is unusual in that the normal to 
the spatial slice is {\it not} perpendicular to the normal of the 
timelike boundary. Because of this, the mass formula is different from 
the standard ADM integral in that it involves the extrinsic curvature 
\cite{hh}. The above approach provides a direct way of obtaining 
flat slice data for the Schwarzschild metric. 

From (\ref{flsch}) it is evident that even with matter, flat slice
data  must have $\tilde{\pi}^{ab}$ falling off at least
as fast as $r^{-3/2}$ to maintain asymptotic flatness. Furthermore the
positive energy condition $\rho>0$ requires from (\ref{asymH}) that
\be 
(f+g)(f-3g)<0
\label{condH}
\ee 
for all $r$. It is difficult to find analytical data explicitly that 
satisfies both these conditions, and (\ref{asymdiff}), without patching 
to Schwarzschild exterior. Consider for example the electric monopole. 
 The solution of $\p_a\tilde{E}^a=0$ is $\tilde{E}^a= Qn^a/r^2$ (with 
the point $r=0$ removed), 
and the magnetic field $F_{ab}=0$. Then the condition on $f(r)$ and 
$g(r)$ from the diffeomorphism constraint is (\ref{asymdiff}), and that 
from the Hamiltonian constraint is 
\be 
f^2 - 2fg - 3g^2 +  {Q^2\over r^4}  = 0. 
\ee 
These two conditions give a single equation for $f(r)$ (or $g(r)$): 
\be 
 f' + {3f\over r} + {\epsilon\over 2} X' = 0,
\ee
where $\epsilon = \pm 1$ and $X = \sqrt{4f^2 + 3E^2 }$, ($E=|E^a|$). 
The same equation applies to scalar field coupling or other matter 
coupling with the appropriate replacement of energy density. 

It is possible to find solutions of this equation numerically. 
The large $r$ behaviour is $\sim r^{-3/2}$ as expected, if the  
matter falls faster than $r^{-3/2}$  at large $r$. For $\epsilon= +1$, 
$f$ and $g$ are positive for large $r$, so the mass of the data 
is positive, and there is an apparent horizon. For $\epsilon=-1$ 
on the other hand, $f$, $g$, and hence the mass are all negative, 
and there is no apparent horizon. This is surprising because the 
matter satsifies the dominant energy condition. It suggests 
a naked singularity like the negative mass Schwarzschild 
case. 

It is easier to find analytic solutions, with matter
satisfying $\rho >0$, if the data is taken to have Schwarzschild 
exterior. An example is provided by taking the interior and exterior
$\tilde{\pi}^{ab}$ given by $g=\alpha + \beta r$, (which implies
$f=-\beta r/3$ (\ref{asymdiff})), and (\ref{flsch}) respectively.
Matching $\tilde{\pi}^{ab}$ at $r=1$ gives
\be 
\alpha = 10 \sqrt{{M\over 2}}, \ \ \ \ \ \ \beta = -9\sqrt{{M\over 2}},
\ee
and interior energy density 
\be 
\rho = 15 M (1-r)(5-3r),\ \ \ \ \ \ r\le 1.
\ee
This energy density may be interpreted as arising from a scalar field 
by setting $\rho = \tilde{P}^2$ and $\phi=0$, or from electromagnetism 
by setting $\rho = \tilde{E}^a\tilde{E}^{b}\delta_{ab}$, $\tilde{B}^a=0$, 
with the Gauss law solved as in the last section. 

The apparent horizon equation for (\ref{gf}) reduces to 
$f(r)+g(r) = 2/r$, and for this example gives 
\be 
r_{AH} = {5\over 6}\left( 1\pm \sqrt{ 1 - {12\over 25}\sqrt{{2\over M}} }
\right)
\label{rah}
\ee
This shows that an apparent horizon forms for $M=2(12/25)^2$ with 
non-zero radius $r=5/6$.  Using (\ref{rah}), it is possible to show 
that Penrose's inequality is satisfied for all $M$. Other examples, 
with $f$ and $g$ polynomials also satisfy the inequality. 

If the condition (\ref{asymdiff}) relating $f(r)$ and $g(r)$ is not 
imposed, then $J^a\ne 0$; the only restrictions on these functions 
are now those arising from energy conditions. An example of 
flat slice asymptotically flat data with $J^a\ne 0$ is 
obtained by setting   
\be 
f(r) = {\alpha\over r^2}, \ \ \ \ \ \ \ g(r) = -{\beta\over r^2}. 
\ee
Then the diffeomorphism and Hamiltonian constraints give  
\be 
J^a = n^a {2\beta \over r^3},\ \ \ \ \ \ \ 
\rho = {(\alpha + 3\beta)(\beta - \alpha)\over 2 r^4}
\ee
respectively. This data satisfies the weak energy condition for all 
$r$ if $\alpha$ and $\beta$ are such that $\rho>0$, but 
violates the dominant energy condition $\rho\ge (J^aJ_a)^{1/2}$ 
for $r > (\alpha + 3\beta)(\beta -\alpha)/4\beta$. The mass of the 
data is zero because the large $r$ falloff of $\tilde{\pi}^{ab}$ is 
faster than the $r^{-3/2}$ required to get a non-zero mass 
from the relevant surface integral \cite{hh}. (Note that it {\it is} 
possible to get a zero mass even if there is a non-zero energy density, 
provided the energy density gives a metric whose leading order behavior 
is $q_{ab} \sim \delta_{ab} + O(1/r^2)$; ie. there is no $1/r$ term 
to be captured by the surface integral. Conversely,  
a form of $q_{ab}$ and $\tilde{\pi}^{ab}$ such that the ADM mass 
is manifestly zero, may be used to deduce an energy density and 
matter current. This is exactly what is done above for flat slice 
data. Although this seems counterintuitive, it is possible: a  
metric with zero ADM mass but non-zero energy density is the 
Reissner-Nordstrom metric with $M=0$ and electric charge $Q\ne 0$; 
while there are no horizons in this case, there are for the above 
flat slice example.) 

The apparent horizon equation is $\alpha - \beta=2r$, showing that 
there are no horizons for $\beta\ge\alpha\ge 0$.  Therefore Penrose's
inequality holds even though the dominant energy condition does
not.  However there are other examples where this is not the case. 
Consider e.g.  $\beta = -\alpha < 0$, for which $\rho = 2\alpha^2/r^4$; 
there is a horizon at $r=\alpha$. Thus,  the inequality can be 
violated if only the weak energy condition holds. 

More generally, from (\ref{gf}) and (\ref{asymH}), the energy 
condition $\rho\ge (J^aJ_a)^{1/2}$ is 
\be 
 {1\over 2}(f+g)(3g - f) = c \left( f' + {2f\over r} 
+ g' \right)\ge 0, 
\ee
for constant $c\ge 1$. Simple ansatze such as $f\sim g$, or $f=g h$
for some function $h$, lead for several analytically solvable cases,
to data which is not asymptotically flat. However, for any given
asymtotically flat form for $f$ or $g$, it is relatively
straightforward to study the resulting ordinary differential equation
numerically to probe Penrose's inequality.
 
\section{Summary}

We have given a number of ansatze for solving the initial value
constraints of general relativity with matter couplings. These include
the massless scalar field, electromagnetic and Yang-Mills fields. Large 
classes of solutions are obtained in each case, both for time-symmetric 
and asymmetric situations. For the scalar field case, explicit 
initial data sets are given for pulses. These may be used as staring 
points for numerical integration. For electromagnetic and Yang-Mills 
fields, the data may be interpreted as the initial geometry due to flux 
lines of the electric field, because of the way the Gauss law is solved 
in each of these cases. These results go beyond the multi-point sources 
considered in earlier works.

For the time-asymmetric cases considered, Penrose's inequality holds
if matter satisfies the dominant energy condition. However, violations
of the inequality can occur if only the weak energy condition holds.

The solutions given in this paper may be used as a possible starting
point for extensions away from spherical symmetry. This would be
useful not just for the data that can be obtained, but also for
providing new tests of cosmic censorship via Penrose's
inequality. This is a potentially important direction because almost
all results pertaining to cosmic censorship are in spherical symmetry.

Several generalizations of the ansatze given here arise if a fixed
direction $S^a$ is specified. Then $\tilde{\pi}^{ab}$ can be
constructed out of $S^a$, $n^a$, and the metric. One may even take
$S^a$ to be a non-constant divergence free vector field specified
similarly to the electric field which solves Gauss's law. These and
similar extensions are presently being studied.

\bigskip

\noindent {\it Acknowledgement:} I would like to thank Ted Jacobson 
and Bill Unruh for discussions. This work was supported by the 
Natural Science and Engineering Research Council of Canada. 

\medskip
\noindent {\it Note added:}  After this work was submitted and posted, 
I learned that Penrose's inequality has been previously studied in 
spherical symmetry: it was proved under certain conditions in 
\cite{mm}, and more generally in \cite{hay}. I thank E. Malec and 
S. Hayward for bringing these references to my attention.

\end{document}